\documentclass[12pt]{article}
\usepackage{amssymb}
\usepackage{theorem}
\newtheorem{theor}{Theorem}

\usepackage{hhline}
\begin{document}
\title{Mean field solutions to singlet hopping and
       superconducting pairing within a two-band Hubbard model}
\author{S.~Adam (1 and 2), Gh.~Adam (1 and 2)\\
((1) LIT-JINR Dubna Russia,\\
(2) IFIN-HH Magurele--Bucharest Romania)}
\maketitle
\begin{abstract}
The mean field Green function solution of the two-band singlet-hole Hubbard
model for high-$T\sb{c}$ superconductivity in cuprates (Plakida, N.M. et al.,
Phys.~Rev. B51, 16599 (1995), JETP 97, 331 (2003))
involves expressions of higher order correlation functions describing
respectively the singlet hopping and the superconducting pairing.
Rigorous derivation of their values is reported based on the finding that
specific invariant classes of polynomial Green functions in terms
of the Wannier overlap coefficients $\nu\sb{ij}$ exist.
\end{abstract}
\section{Introduction}
The two-band singlet-hole Hubbard model considered by Plakida et al.~%
\cite{Pl95}
for the description of the high-$T\sb{c}$ superconductivity in cuprates in
terms of Hubbard operators (HOs) provides the simplest consistent approach
towards the incorporation of the essential features of these systems (strong
antiferromagnetic superexchange interaction inside the $CuO\sb{2}$ planes,
occurrence of two relatively isolated energy bands around the Fermi
level, able to develop $d\sb{x\sp{2}-y\sp{2}}$ pairing) such as to describe
simultaneously both the normal and the superconducting states within a frame
which secures rigorous fulfilment of the basic principles of the quantum
mechanics.
The equation of motion method for two-time Green functions
\cite{Zub60}
was successfully used to derive electron spectra of the model
\cite{Pl95}
and to incorporate the superconducting state as well
\cite{Pl97,Pl03}.
\par
The present paper reports rigorous results for the expressions of two
higher order correlation functions which arise in the generalized mean
field approximation (GMFA) solution of the Green function (GF)
\cite{Pl03}: $\langle X\sb{i}\sp{02}X\sb{j}\sp{20}\rangle$ which describes
the \emph{singlet hopping}, and 
$\langle X\sb{i}\sp{02}N\sb{j}\rangle$, which describes the \emph{exchange
superconducting pairing interaction}.
\par
The idea proposed in 
\cite{Pl03}
for the evaluation of the average
  $\langle X\sb{i}\sp{02}N\sb{j}\rangle$
can be consistently generalized to yield power series expansions of both
statistical averages.
We find that the lowest order expressions of the two correlation functions,
generically denoted henceforth
  $\langle X\sb{i}\sp{02}Q\sb{j}\rangle$
$-$ where $Q\sb{j}$ is either $X\sb{j}\sp{20}$ or $N\sb{j}$, are
\emph{obtained in terms of GMFA Green functions}.
\par
This remarkable result follows from the mathematical properties of the Hubbard
operators. These allow the definition of characteristic \emph{invariant
classes of polynomial Green functions\/} in terms of the Wannier overlap
coefficients
$\nu\sb{ij}$, which are characterized by the property that, under the
iteration of the equation of motion, the operator part remains
\emph{invariant\/}, while the polynomial degree in $\nu\sb{ij}$ is
increased making thus possible consistent power series expansions in the
small parameters $\nu\sb{ij}$.
\section{Statement of the problem}
{\bf 2.1. Hubbard operators.}
The Hubbard operators (HOs)
$X\sb{i}\sp{\alpha \beta} = |i\alpha\rangle \langle i\beta|$
are defined for the four states of the model:
$|0\rangle$ (vacuum), $|\sigma\rangle = \vert \! \! \uparrow \rangle$ and
$|\bar\sigma\rangle = \vert \! \! \downarrow \rangle$ (spin states inside
the hole subband; in numerical calculations, 
$\sigma = \pm 1/2, \bar{\sigma} = -\sigma$), and
$|2\rangle = \vert \! \! \uparrow \downarrow \rangle$ (singlet state in the
singlet subband).
\par
The multiplication rule holds
  $X\sb{i}\sp{\alpha \beta}X\sb{i}\sp{\gamma\eta} = \delta \sb{\beta\gamma}
   X\sb{i}\sp{\alpha\eta}$.
The HOs describing the creation/de\-struc\-tion of single states in a
subband are Fermi-like ones and obey the anticommutation relations
  $\{X\sb{i}\sp{\alpha \beta}, X\sb{j}\sp{\gamma\eta}\} = \delta\sb{ij}
           (\delta \sb{\beta\gamma}X\sb{i}\sp{\alpha\eta} +
            \delta \sb{\eta\alpha}X\sb{i}\sp{\gamma\beta})$,
while the HOs describing the creation/destruction of singlets, spin and
charge densities, particle numbers, are Bose-like ones and obey the
commutation relations
  $[X\sb{i}\sp{\alpha \beta}, X\sb{j}\sp{\gamma\eta}] =
   \delta\sb{ij}(\delta \sb{\beta\gamma}X\sb{i}\sp{\alpha\eta} -
   \delta \sb{\eta\alpha}X\sb{i}\sp{\gamma\beta})$.
At each lattice site $i$, the constraint of no double occupancy of any
quantum state $|i\alpha\rangle$ is rigorously preserved due to the
completeness relation
  $X\sb{i}\sp{00} + X\sb{i}\sp{\sigma\sigma} +
   X\sb{i}\sp{\bar{\sigma}\bar{\sigma}} + X\sb{i}\sp{22} = 1$.
\par
The \emph{particle number operator at site $i$} is given by
\begin{equation}
    N\sb{i} = X\sb{i}\sp{\sigma\sigma}+
              X\sb{i}\sp{\bar\sigma\bar\sigma}+2X\sb{i}\sp{22}.
\label{eq:N}
\end{equation}
\par
We define the \emph{Hubbard p-form of labels\/} $(\alpha\beta,\gamma\eta)$,
\begin{equation}
  \tau\sp{\alpha\beta,\gamma\eta}\sb{p,i} =
  \sum\sb{j\neq i}\nu\sb{ij}\sp{p}X\sb{i}\sp{\alpha \beta}
  X\sb{j}\sp{\gamma\eta} , \quad p=1, 2, \cdots ,
\label{eq:taup}
\end{equation}
where the meanings of HOs $X\sb{i}\sp{\alpha \beta}$ and
$X\sb{j}\sp{\gamma\eta}$ depend on the context.
The Wannier overlap coefficients $\nu\sb{ij}$ are small quantities rapidly
decreasing with the 
intersite distance $r\sb{ij}=|{\bf r}\sb{j}-{\bf r}\sb{i}|$ (see, e.g.,
\cite{FJR96,YOH97}
and references quoted therein).
The nearest neighbour values
$\nu\sb{i, i\pm a\sb{x/y}} = \nu\sb{1} \simeq -0.14$
and next nearest ones
$\nu\sb{i, i \pm a\sb{x} \pm a\sb{y}} = \nu\sb{2} \simeq -0.02$ considered in
\cite{Pl95}
are typical.
\par
{\bf 2.2. Model Hamiltonian.}
The Hamiltonian of the model
\cite{Pl95}
can be rewritten in terms of linear Hubbard forms describing hopping processes
as follows
\begin{eqnarray}
  \! \! \! &&\! \! \! H =
    E\sb{1} \sum\sb{i,\sigma} X\sb{i}\sp{\sigma \sigma} \! +\!
      E\sb{2} \sum\sb{i} X\sb{i}\sp{22} \! +\!
\nonumber\\
  \! \! \! &&\quad \quad +
 \mathcal{K}\sb{11}\sum\sb{i,\sigma} \tau\sb{1,i}\sp{\sigma 0, 0 \sigma} \! +\!
 \mathcal{K}\sb{22}\sum\sb{i,\sigma} \tau\sb{1,i}\sp{2\sigma , \sigma 2} \! +\!
   \mathcal{K}\sb{12}\sum\sb{i,\sigma}2\sigma
           (\tau\sb{1,i}\sp{2\bar{\sigma},0\sigma} \! +\!
            \tau\sb{1,i}\sp{\sigma 0, \bar{\sigma}2}),
\label{eq:H}
\end{eqnarray}
where the summation label $i$ runs over the sites of
an infinite two-dimensional square array the lattice constants
of which, $a\sb{x} = a\sb{y}$, are defined by the underlying single
crystal structure.
\par
In Eq.~(\ref{eq:H}), $E\sb{1} = \tilde{\varepsilon \sb{d}} - \mu$ and
$E\sb{2} = 2E\sb{1} + U\sb{eff}$, where $\tilde{\varepsilon \sb{d}}$ is
the renormalized energy of a $d$-hole, $\mu$ is the chemical potential, while
$U\sb{eff} \equiv \Delta \approx \Delta \sb{pd} =
  \varepsilon \sb{p} - \varepsilon \sb{d}$
is an effective Coulomb energy corresponding to the difference between
the hole energy levels for oxygen and copper.
\par
Keeping in mind that the lower label $1$ refers to one-hole states, while
the lower label $2$ to singlet states, the quantities
$\mathcal{K}\sb{ab} = 2 t\sb{pd}K\sb{ab}$ are characteristic hopping energies
for either inband ($a=b$) or interband ($a\neq b;
\mathcal{K}\sb{12}=\mathcal{K}\sb{21}$) transitions between the
two bands of the model. Here $t\sb{pd}$ denotes the hopping $p$-$d$ integral
and $K\sb{ab}$ are numerical coefficients coming from hybridization
effects between the holes and the singlets
\cite{Pl95}.
\par
The translational invariance of the system gives
\begin{equation}
    (\tau\sb{i}\sp{\alpha \beta, \gamma \eta})\sp{\dagger} =
   - \tau\sb{i}\sp{\beta \alpha, \eta \gamma} =
    \tau\sb{i}\sp{\eta \gamma, \beta \alpha},
\label{eq:taudag}
\end{equation}
which secures the hermiticity of the model Hamiltonian $H$.
\par
{\bf 2.3. Mean field approximation.}
The quasi-particle spectrum and superconducting pairing within the model
Hamiltonian~(\ref{eq:H}) are obtained
\cite{Pl97,Pl03}
from the two-time $4\times 4$ matrix Green function (GF) in Zubarev
notation
\cite{Zub60}
\begin{equation}
         \tilde G\sb{ij\sigma}(t-t')  =
    \langle\langle \hat X\sb{i\sigma}(t)\! \mid \!
    \hat X\sb{j\sigma}\sp{\dagger}(t')\rangle\rangle =
    -i\theta (t-t')\langle \{\hat X\sb{i\sigma}(t),
    \hat X\sb{j\sigma}\sp{\dagger}\}\rangle .
\label{eq:GF}
\end{equation}
where $\langle \cdots \rangle$ denotes
the statistical average over the Gibbs grand canonical ensemble.
\par
The GF~(\ref{eq:GF}) is defined for the four-component Nambu column operator
\begin{equation}
  \hat X\sb{i\sigma}=(X\sb{i}\sp{\sigma 2}\,\,
  X\sb{i}\sp{0\bar\sigma}\,\, X\sb{i}\sp{2\bar\sigma}\,\,
  X\sb{i}\sp{\sigma 0})\sp{({\rm T})}
\label{eq:nambu}
\end{equation}
The operator
  $\hat X\sb{j\sigma}\sp{\dagger} = (X\sb{j}\sp{2\sigma}\,\,
  X\sb{j}\sp{\bar\sigma 0}\,\, X\sb{j}\sp{\bar\sigma 2}\,\,
  X\sb{j}\sp{0\sigma})$
is the adjoint of $\hat X\sb{j\sigma}$.
In (\ref{eq:nambu}), the superscript (T) denotes the transposition.
Here and in what follows, due to translational invariance,
the notation $\mathcal{G}\sb{ij}$ points to the
dependence of the quantity $\mathcal{G}$ of interest on the
\emph{distance} $r\sb{ij} = |{\bf r}\sb{j} - {\bf r}\sb{i}|$
between the position vectors of the lattice sites $j$ and $i$ respectively.
\par
The derivation of the GF within GMFA needs the knowledge of the
\emph{frequency matrix},
\begin{equation}
  \tilde \mathcal{A}\sb{ij\sigma} = \langle \{ [\hat X\sb{i\sigma}, H],
    \hat X\sb{j\sigma}\sp{\dagger} \} \rangle .
\label{eq:Aij}
\end{equation}
\par
Direct calculations
\cite{Pl03}
show that the \emph{normal\/} matrix elements of
$\tilde \mathcal{A}\sb{ij\sigma}$ contain \emph{one-site} $(i=j)$ GMFA
hopping correlation function which result in 
$\mathcal{O}(\mathcal{K}\sb{ab}\nu\sb{ij})$ renormalizations of
the energy parameters $E\sb{1}$ and $E\sb{2}$, as well as \emph{two-site}
$(i\neq j)$ hopping generated \emph{higher order\/} correlation functions
bringing two distinct kinds of contributions to
$\tilde \mathcal{A}\sb{ij\sigma}$: \emph{charge-spin} correlations
(which can be conveniently subdivided into \emph{charge-charge},
$\langle N\sb{i}N\sb{j}\rangle$, and \emph{spin-spin},
$\langle{\bf S}\sb{i}{\bf S}\sb{j}\rangle =
 \langle X\sb{i}\sp{\sigma\bar\sigma}X\sb{j}\sp{\bar\sigma\sigma}\rangle$,
correlations) and the \emph{singlet hopping} correlation
function $\langle X\sb{i}\sp{02}X\sb{j}\sp{20}\rangle$ (singlet destruction
at site $i$ followed by singlet creation at site $j$).
\par
There are three distinct matrix elements out of the eight normal matrix
elements of $\tilde \mathcal{A}\sb{ij\sigma}$ containing
$\langle X\sb{i}\sp{02}X\sb{j}\sp{20}\rangle$:
\begin{eqnarray}
  && (\sigma 2, 2\sigma ) \quad\quad\quad -\mathcal{K}\sb{11}\nu\sb{ij}
       \langle X\sb{i}\sp{02}X\sb{j}\sp{20}\rangle
\label{eq:sh22}\\
  && (0\bar\sigma , \bar\sigma 0) \quad\quad\quad -\mathcal{K}\sb{22}\nu\sb{ij}
       \langle X\sb{i}\sp{02}X\sb{j}\sp{20}\rangle
\label{eq:sh11}\\
  && (\sigma 2, \bar\sigma 0)  \ \; \quad
   -2\sigma\! \cdot \! \mathcal{K}\sb{21}
       \nu\sb{ij}\langle X\sb{i}\sp{02}X\sb{j}\sp{20}\rangle
\label{eq:sh21}
\end{eqnarray}
\par
The only non-vanishing \emph{anomalous\/} matrix elements of
$\tilde \mathcal{A}\sb{ij\sigma}$ are
\cite{Pl03}
the hopping generated \emph{two-site} contributions involving the higher
order correlation function $\langle X\sb{i}\sp{02}N\sb{j}\rangle$.
This provides the \emph{exchange superconducting
pairing mechanism\/} originating in the interaction of an anomalous pair
of particles at a same site $i$ but in different subbands
$(X\sb{i}\sp{02} = X\sb{i}\sp{0\sigma}X\sb{i}\sp{\sigma2})$, with
the surrounding particle distribution at the neighbouring site $j$
described by the particle number operator $N\sb{j}$, Eq.~(\ref{eq:N}).
The structure of the anomalous part of $\tilde \mathcal{A}\sb{ij\sigma}$
is very special:
\begin{eqnarray}
  && (\sigma 2, \bar\sigma 2) \quad\quad\quad\quad\quad\;\!
       2\bar\sigma\! \cdot \! \mathcal{K}\sb{21}
       \nu\sb{ij}\langle X\sb{i}\sp{02}N\sb{j}\rangle
\label{eq:aav22}\\
  && (0\bar\sigma , \sigma 0) \quad\quad\quad\quad\quad\;\!
       2\sigma\! \cdot \! \mathcal{K}\sb{21}
       \nu\sb{ij}\langle X\sb{i}\sp{02}N\sb{j}\rangle
\label{eq:aav11}\\
  && (\sigma 2, 0\sigma ) \quad\quad \; \frac{1}{2}
      (\mathcal{K}\sb{11}+\mathcal{K}\sb{22})\nu\sb{ij}
       \langle X\sb{i}\sp{02}N\sb{j}\rangle
\label{eq:aav21}\\
  && (0\bar\sigma , \bar\sigma 2) \quad -\frac{1}{2}
      (\mathcal{K}\sb{11}+\mathcal{K}\sb{22})\nu\sb{ij}
       \langle X\sb{i}\sp{02}N\sb{j}\rangle
\label{eq:aav12}
\end{eqnarray}
The other anomalous matrix elements are obtained by complex conjugation.
\section{Fundamental relationships}
From the spectral theorem
\cite{Zub60},
\begin{equation}
  \langle X\sb{i}\sp{02}Q\sb{j}\rangle =
  \frac{i}{2\pi}\int\limits_{-\infty}^{+\infty}
  \frac{{\rm d}\omega}{1-{\rm e}\sp{-\beta\omega}}
  \Big[\langle\langle X\sb{i}\sp{02}|Q\sb{j}\rangle\rangle
       \sb{\omega+i\varepsilon} -
  \langle\langle X\sb{i}\sp{02}|Q\sb{j}\rangle\rangle
       \sb{\omega-i\varepsilon}\Big],
\label{eq:SpT}
\end{equation}
where the labels $\pm i\varepsilon$, $\varepsilon = 0\sp{+}$, refer to the
retarded/advanced Green
functions respectively. Since both $X\sb{i}\sp{02}$ and $Q\sb{j}$ are bosonic
Hubbard operators, the thermodynamic factor in the denominator
is $1-{\rm e}\sp{-\beta\omega}$ and the two Green
functions are defined in terms of the commutators of the two operators, i.e.,
\begin{equation}
  \langle\langle X\sb{i}\sp{02}(t)|Q\sb{j}(t')\rangle\rangle =
    -i\theta(t-t')\langle[X\sb{i}\sp{02}(t), Q\sb{j}(t')]\rangle
\label{eq:GFXQr}
\end{equation}
for the retarded Green function and a similar definition for the advanced one.
\par
By differentiation with respect to $t$ and use of Fourier transform, we
get the following basic result for the two Green functions in the
$({\bf r}, \omega)$-representation required by Eq.~(\ref{eq:SpT})
(for the sake of simplicity, $\pm i\varepsilon$ terms are omitted):
\begin{eqnarray}
   (\omega-E\sb{2})
    \langle\langle X\sb{i}\sp{02}|Q\sb{j}\rangle\rangle\sb{\omega}=
    \! \! \! \! \! &-&\! \! \! \! \!
    \mathcal{K}\sb{11}\sum\sb{\sigma}
    \langle\langle\tau\sb{1,i}\sp{\sigma 2,0\sigma}|
                     Q\sb{j}\rangle\rangle\sb{\omega} +
    \mathcal{K}\sb{22}\sum\sb{\sigma}
    \langle\langle\tau\sb{1,i}\sp{0\sigma,\sigma 2}|
                     Q\sb{j}\rangle\rangle\sb{\omega} +
\nonumber\\
    \! \! \! \! \! &+&\! \! \! \! \!
    \mathcal{K}\sb{21}\sum\sb{\sigma}2\sigma\Big(
    \langle\langle\tau\sb{1,i}\sp{0\bar\sigma ,0\sigma}|
                     Q\sb{j}\rangle\rangle\sb{\omega} -
    \langle\langle\tau\sb{1,i}\sp{\sigma 2,\bar\sigma 2}|
                     Q\sb{j}\rangle\rangle\sb{\omega}\Big)
\label{eq:GFro}
\end{eqnarray}
\begin{theor}
 Let
  $g\sp{\alpha\beta,\gamma\eta}\sb{2p-1}\equiv 
   \langle\langle\tau\sp{\alpha\beta,\gamma\eta}\sb{2p-1,i}\vert
      Q\sb{j}\rangle\rangle\sb{\omega}$
 be a generic notation of the \emph{extensions to Hubbard ($2p-1$)-forms}
 (\ref{eq:taup}) of the four Green functions
 which enter the r.h.s.~of Eq.~(\ref{eq:GFro}).
 Then the recurrence relations hold,
\begin{eqnarray}
   \! \! \! \! \! \! \! g\sb{2p-1}\sp{\sigma 2, 0\sigma}
   \! \! \! \! &=&\! \! \! \!
   \frac{\nu\sb{ij}\sp{2p-1}M'\sb{ij\sigma}}{\omega-E\sb{2}}\! -\!
   \frac{\mathcal{K}\sb{22}\nu\sb{ij}\sp{2p}P''\sb{ij\sigma}}
        {(\omega-E\sb{2})\sp{2}}\! +\!
   \frac{\mathcal{K}\sb{11}\sp{2} + \mathcal{K}\sb{22}\sp{2}}
        {(\omega-E\sb{2})\sp{2}}\, g\sb{2p+1}\sp{\sigma 2, 0\sigma}\! -\!
   \frac{2\mathcal{K}\sb{11} \mathcal{K}\sb{22}}
        {(\omega-E\sb{2})\sp{2}}\, g\sb{2p+1}\sp{0\sigma, \sigma 2}
\label{eq:G21}\\
   \! \! \! \! \! \! \! g\sb{2p-1}\sp{0\sigma, \sigma 2}
   \! \! \! \! &=&\! \! \! \!
   \frac{\nu\sb{ij}\sp{2p-1}M''\sb{ij\sigma}}{\omega-E\sb{2}}\! +\!
   \frac{\mathcal{K}\sb{11}\nu\sb{ij}\sp{2p}P''\sb{ij\sigma}}
        {(\omega-E\sb{2})\sp{2}}\! -\!
   \frac{2\mathcal{K}\sb{11} \mathcal{K}\sb{22}}
        {(\omega-E\sb{2})\sp{2}}\, g\sb{2p+1}\sp{\sigma 2, 0\sigma}\! +\!
   \frac{\mathcal{K}\sb{11}\sp{2} + \mathcal{K}\sb{22}\sp{2}}
        {(\omega-E\sb{2})\sp{2}}\, g\sb{2p+1}\sp{0\sigma, \sigma 2}
\label{eq:G12}\\
   \! \! \! \! \! \! \! g\sb{2p-1}\sp{0\bar\sigma,0\sigma}
   \! \! \! \! &=&\! \! \! \!
   \frac{\nu\sb{ij}\sp{2p-1}M'''\sb{ij\sigma}}{\omega-2E\sb{1}}\! +\!
   \frac{2\sigma\mathcal{K}\sb{21}\nu\sb{ij}\sp{2p}
        (P'''\sb{ij\sigma}\! +\! P\sb{ij\sigma}\sp{IV})}
        {(\omega-2E\sb{1})(\omega-E\sb{2})}\! +\!
   \frac{(2\mathcal{K}\sb{21} \cdot 2\sigma)\sp{2}}
        {(\omega-2E\sb{1})(\omega-E\sb{2})}g\sb{2p+1}\sp{0\bar\sigma,0\sigma}
\label{eq:G11}\\
   \! \! \! \! \! \! \! g\sb{2p-1}\sp{\sigma 2,\bar\sigma 2}
   \! \! \! \! &=&\! \! \! \!
   \frac{\nu\sb{ij}\sp{2p-1}M\sb{ij\sigma}\sp{IV}}
        {\omega\! -\! (E\sb{2}\! \! +\! \! \Delta)}\! +\!
   \frac{2\bar\sigma\mathcal{K}\sb{21}\nu\sb{ij}\sp{2p}
        (P\sb{ij\sigma}\sp{V}\! \! +\! \! P\sb{ij\sigma}\sp{VI})}
        {[\omega\! -\! (E\sb{2}\! \! +\! \! \Delta)]
        (\omega\! -\! E\sb{2})}\! +\!
   \frac{(2\mathcal{K}\sb{21} \cdot 2\bar\sigma)\sp{2}}
        {[\omega\! -\! (E\sb{2}\! \! +\! \! \Delta)](\omega\! -\! E\sb{2})}
          g\sb{2p+1}\sp{\sigma 2,\bar\sigma 2}
\label{eq:G22}
\end{eqnarray}
where the coefficients $M\sb{ij\sigma}$ and $P\sb{ij\sigma}$, given in
Table~1, are statistical averages following from equal time commutator terms.
\label{theor:tauQ}
\end{theor}
The proof is immediate if we write the equations of motion of the Green
functions mentioned in the l.h.s. and iterate once.
 \begin{table}[ht]
   \caption{
Equal time commutators arising in the recurrence relations
(\ref{eq:G21})--(\ref{eq:G22}) as coefficients of
$\nu\sb{ij}\sp{2p-1}$ and $\nu\sb{ij}\sp{2p}$
           }
   \label{tab:coeff}
  \begin{center}
  \renewcommand{\arraystretch}{1.5}
  \setlength{\doublerulesep}{.3pt}
   \begin{tabular}{||c|c|c||c|c|c||}
    \hhline{|t:===:t:===:t|}
        $\nu\sb{ij}\sp{2p-1}$ &
   $Q\sb{j} = X\sb{j}\sp{20}$ & $Q\sb{j} = N\sb{j}$ &
        $\nu\sb{ij}\sp{2p}$   &
   $Q\sb{j} = X\sb{j}\sp{20}$ & $Q\sb{j} = N\sb{j}$\\
    \hhline{#===#===#}
            $M'\sb{ij\sigma}$ &
   $-\langle X\sb{i}\sp{\sigma 2}X\sb{j}\sp{2\sigma}\rangle$    &
   $\langle X\sb{i}\sp{\sigma 2}X\sb{j}\sp{0\sigma}\rangle$     &
            $P'\sb{ij\sigma}$ &
     0                        &  0                             \\
    \hline
           $M''\sb{ij\sigma}$ &
    $\langle X\sb{i}\sp{0\sigma}X\sb{j}\sp{\sigma 0}\rangle$    &
    $\langle X\sb{i}\sp{0\sigma}X\sb{j}\sp{\sigma 2}\rangle$    &
           $P''\sb{ij\sigma}$ &
    $\langle X\sb{i}\sp{\sigma\sigma}(X\sb{j}\sp{00}-X\sb{j}\sp{22})\rangle$ &
    $\langle X\sb{i}\sp{\sigma\sigma}X\sb{j}\sp{02}\rangle$ \\
    \hhline{#===#===#}
          $M'''\sb{ij\sigma}$ &
 $-\langle X\sb{i}\sp{0\bar\sigma}X\sb{j}\sp{2\sigma}\rangle$   &
 $\langle X\sb{i}\sp{0\bar\sigma}X\sb{j}\sp{0\sigma}\rangle$    &
          $P'''\sb{ij\sigma}$ &
 $-\langle X\sb{i}\sp{02}X\sb{j}\sp{20}\rangle$                 &
 $                   0                                     $   \\
    \hline
      $M\sb{ij\sigma}\sp{IV}$ &
 $\langle X\sb{i}\sp{\sigma 2}X\sb{j}\sp{\bar\sigma 0}\rangle$  &
 $\langle X\sb{i}\sp{\sigma 2}X\sb{j}\sp{\bar\sigma 2}\rangle$  &
      $P\sb{ij\sigma}\sp{IV}$ &
 $\langle X\sb{i}\sp{00}(X\sb{j}\sp{00}-X\sb{j}\sp{22})\rangle$ &
 $2\langle X\sb{i}\sp{00}X\sb{j}\sp{02}\rangle$ \\
    \hhline{#===#===#}
                              &
                              &
                              &
       $P\sb{ij\sigma}\sp{V}$ &
 $\langle X\sb{i}\sp{02}X\sb{j}\sp{20}\rangle$                  &
 $                   0                                     $   \\
    \hline
                              &
                              &
                              &
      $P\sb{ij\sigma}\sp{VI}$ &
 $\langle X\sb{i}\sp{22}(X\sb{j}\sp{00}-X\sb{j}\sp{22})\rangle$ &
 $2\langle X\sb{i}\sp{22}X\sb{j}\sp{02}\rangle$                \\
    \hhline{|b:===:b:===:b|}
   \end{tabular}
  \end{center}
 \end{table}
\par
  The \emph{$p$-form class invariance} of the abovementioned Green functions
is a straightforward consequence of the commutation and multiplication
rules satisfied by the Hubbard operators.
It allows the derivation of the statistical averages
 $\langle X\sb{i}\sp{02}X\sb{j}\sp{20}\rangle$ and
 $\langle X\sb{i}\sp{02}N\sb{j}\rangle$ as power series
of the small parameters of the model $\mathcal{K}\sb{11}\nu\sb{ij}$,
$\mathcal{K}\sb{22}\nu\sb{ij}$, and $\mathcal{K}\sb{21}\nu\sb{ij}$.
\par
  While the coefficients of the odd powers of the series expansions are
obtained in terms of GMFA Green functions, those of the even powers are
still defined in terms of Green functions beyond GMFA.
From the point of view of the practical implications, it is an academic
question whether it would be possible to express them in terms of GMFA
Green functions as well. The lowest order approximations provide the most
important contributions to the observables.
The question is whether the corresponding statistical averages quoted in the
Table~1 are really significant or not. This point is discussed in the
next section.
\section{Significant lowest order terms}
The results derived in the previous section show that the statistical averages
$\langle X\sb{i}\sp{02}Q\sb{j}\rangle$ are obtained as power series of
$\nu\sb{ij}$, with the contributions coming from the poles of the Green
functions given by integrals of the form
\begin{eqnarray}
  I\sb{mn}(\omega\sb{1}, \omega\sb{2})\! \! &=&\! \!
  I\sb{mn}\sp{-}(\omega\sb{1}, \omega\sb{2}) -
  I\sb{mn}\sp{+}(\omega\sb{1}, \omega\sb{2}),\quad m+n\geq 2, \quad m, n>0,
\label{eq:Imn}\\
  I\sb{mn}\sp{\mp}(\omega\sb{1}, \omega\sb{2})\! \! &=&\! \!
  \frac{i}{2\pi}\int\limits_{-\infty}^{+\infty}
  \frac{{\rm d}\omega}{1-{\rm e}\sp{-\beta\omega}}
  \frac{1}{[\omega\! -\! (\omega\sb{1}\mp i\varepsilon)]\sp{m}}\cdot
  \frac{1}{[\omega\! -\! (\omega\sb{2}\mp i\varepsilon)]\sp{n}}
\label{eq:Imnpm}
\end{eqnarray}
The calculation of these integrals is standard: they are extended in
corresponding complex half-planes $z=(\omega, \Im z\lessgtr 0)$,
with half-circles at the three existing poles: $z=0$,
$z=\omega\sb{1}\pm i\varepsilon$, $z=\omega\sb{2}\pm i\varepsilon$.
The obtained contour integrals are calculated in two alternatives ways:
using the residue theorem, and estimating them along the pieces of the
involved contours.
\par
Retaining the lowest order contributions to
 $\langle X\sb{i}\sp{02}Q\sb{j}\rangle$
only, we get:
\begin{eqnarray}
\langle X\sb{i}\sp{02}Q\sb{j}\rangle \! \! \! \! \! &=&\! \! \! \! \!
  I\sb{20}\! (E\sb{2})c\sb{1}\sp{\sigma 2,0\sigma}\! \! +\!
  I\sb{11}\! (E\sb{2},\! 2E\sb{1})c\sb{1}\sp{0\bar\sigma,0\sigma}\! \! +\!
  I\sb{11}\! (E\sb{2}\! \! +\! \! \Delta, E\sb{2})c\sb{1}\sp{\sigma 2,\bar\sigma 2}
      \! \! +\! \! \mathcal{O}(\nu\sb{ij}\sp{2}),
\nonumber\\
  c\sb{1}\sp{\sigma 2,0\sigma}\! \! \! \! \! &=&\! \! \! \! \!
   -(2\mathcal{K}\sb{11}\nu\sb{ij})M'\sb{ij\sigma}\! \! +\! \!
   (2\mathcal{K}\sb{22}\nu\sb{ij})M''\sb{ij\sigma},
\nonumber\\
  c\sb{1}\sp{0\bar\sigma,0\sigma}\! \! \! \! \! &=&\! \! \! \! \!
   2\sigma(2\mathcal{K}\sb{21}\nu\sb{ij})M'''\sb{ij\sigma},
\nonumber\\
  c\sb{1}\sp{\sigma 2,\bar\sigma 2}\! \! \! \! \! &=&\! \! \! \! \!
   2\bar\sigma(2\mathcal{K}\sb{21}\nu\sb{ij})M\sb{ij\sigma}\sp{IV},
\nonumber\\
  I\sb{20}(E\sb{2})\! \! \! \! \! &=&\! \! \! \! \!
 \frac{1}{\beta E\sb{2}\sp{2}}\! -\!
 \frac{\beta {\rm e}\sp{-\beta E\sb{2}}}{(1-{\rm e}\sp{-\beta E\sb{2}})\sp{2}}
\nonumber\\
  I\sb{11}\! (E\sb{2},\! 2E\sb{1})\! \! \! \! \! &=&\! \! \! \! \!
  \frac{1}{2\beta E\sb{1}E\sb{2}}\! -\! 
  \frac{1}{\Delta}\! \cdot \! \frac{1}{1-{\rm e}\sp{-2\beta E\sb{1}}}\! +\!
  \frac{1}{\Delta}\! \cdot \! \frac{1}{1-{\rm e}\sp{-\beta E\sb{2}}}
\nonumber\\
  I\sb{11}\! (E\sb{2}\! \! +\! \! \Delta,\! E\sb{2})
      \! \! \! \! \! &=&\! \! \! \! \!
  \frac{1}{\beta E\sb{2}(E\sb{2}\! \! +\! \! \Delta)}\! +\! 
  \frac{1}{\Delta}\! \cdot \! \frac{1}{1-{\rm e}\sp{-\beta (E\sb{2} + \Delta)}}\! -\!
  \frac{1}{\Delta}\! \cdot \! \frac{1}{1-{\rm e}\sp{-\beta E\sb{2}}}.
\label{eq:Xi02Qj}
\end{eqnarray}
\begin{theor}
 The lowest order power series expansions of the correlation functions
$\langle X\sb{i}\sp{02}X\sb{j}\sp{20}\rangle$ and
$\langle X\sb{i}\sp{02}N\sb{j}\rangle$ are obtained in terms of GMFA
correlation functions as follows:
\par
For \emph{hole-doped\/} systems:
\begin{equation}
  \langle X\sb{i}\sp{02}X\sb{j}\sp{20}\rangle \simeq
  2\bar\sigma\frac{\mathcal{K}\sb{21}\nu\sb{ij}}{\Delta}
  \langle X\sb{i}\sp{\sigma 2}X\sb{j}\sp{\bar\sigma 0}\rangle ;\quad
  \langle X\sb{i}\sp{02}N\sb{j}\rangle \simeq
  2\bar\sigma\frac{\mathcal{K}\sb{21}\nu\sb{ij}}{\Delta}
  \langle X\sb{i}\sp{\sigma 2}X\sb{j}\sp{\bar\sigma 2}\rangle .
\label{eq:hd}
\end{equation}
\par
For \emph{electron-doped\/} systems:
\begin{equation}
  \langle X\sb{i}\sp{02}X\sb{j}\sp{20}\rangle \simeq
  2\bar\sigma\frac{\mathcal{K}\sb{21}\nu\sb{ij}}{\Delta}
  \langle X\sb{i}\sp{0\bar\sigma}X\sb{j}\sp{2\sigma}\rangle ;\quad
  \langle X\sb{i}\sp{02}N\sb{j}\rangle \simeq
  2\sigma\frac{\mathcal{K}\sb{21}\nu\sb{ij}}{\Delta}
  \langle X\sb{i}\sp{0\bar\sigma}X\sb{j}\sp{0\sigma}\rangle .
\label{eq:ed}
\end{equation}

\label{theor:lord}
\end{theor}
Indeed, for hole-doped systems, the Fermi level lays in the upper \emph{singlet
subband}, such that we get the energy parameter estimates
 $E\sb{1}\simeq E\sb{2}\simeq -\Delta$. This yields
$I\sb{20}(E\sb{2})\simeq I\sb{11}(E\sb{2},2E\sb{1})\simeq 0$, while
$I\sb{11}(E\sb{2}+\Delta,E\sb{2})\simeq (2\Delta)\sp{-1}$, therefrom
Eqs.~(\ref{eq:hd}) follow.
\par
For electron-doped systems, the Fermi level lays in the upper \emph{hole
subband}, and this yields the energy parameter estimates
$E\sb{1}\simeq 0, \ E\sb{2}\simeq \Delta$. This results in
$I\sb{20}(E\sb{2})\simeq I\sb{11}(E\sb{2}+\Delta,E\sb{2})\simeq 0$, while
$I\sb{11}(E\sb{2},2E\sb{1})\simeq (2\Delta)\sp{-1}$, hence Eqs.~(\ref{eq:ed})
follow.
\section{Discussion of the results}
Corroboration of Eqs.~(\ref{eq:sh22}) to~(\ref{eq:aav12}) with the results
stated in Theorem~\ref{theor:lord} show that, both in hole-doped and
electron-doped cuprates, the dominant contributions to the
singlet hopping and the exchange superconducting pairing are second order
effects described by GMFA correlation functions.
In~(\ref{eq:hd}) and~(\ref{eq:ed}) there occurs a \emph{same}
small parameter, $\mathcal{K}\sb{21}\nu\sb{ij}/{\Delta}$, for the description
of all the involved higher order correlation functions with, however,
\emph{specific} GMFA correlation functions. Thus, the singlet
hopping proceeds by~$i\rightleftarrows j$ jumps of a particle from the
upper energy subband to the lower energy subband. The anomalous
superconducting pairing involves two spin states at neighbouring lattice
sites $i$ and~$j$, both with energies in that subband which crosses the
Fermi level.
However, while both processes are given by small
$\mathcal{O}(\nu\sb{ij}\sp{2})$ quantities, their consequences are quite
different.
\par
The non-vanishing singlet hopping brings a small correction to
the energy terms entering the normal part of
$\tilde \mathcal{A}\sb{ij\sigma}$. Therefore, the decoupling ansatz
$\langle X\sb{i}\sp{02}X\sb{j}\sp{20}\rangle \approx
   \langle X\sb{i}\sp{02}\rangle\langle X\sb{j}\sp{20}\rangle = 0$
used in
\cite{Pl95}
does not substantially modify the general picture obtained for the normal
state.
\par
On the other side, the derivation of the correct GMFA contribution to
$\langle X\sb{i}\sp{02}N\sb{j}\rangle$ is essential for understanding the
pairing mechanism emerging within the model.
Under the assumption of uniform hopping,
$K\sb{11}=K\sb{22}=K\sb{21}=K$, Eqs.~(\ref{eq:aav22})--(\ref{eq:aav12})
and~(\ref{eq:hd})--(\ref{eq:ed}) result in the
well-known AFM exchange interaction energy of the $t$-$J$ model,
$J=4t\sp{2}/\Delta$, for spins on nearest neighbour sites and an effective
hopping parameter $t=t\sb{pd}K\nu\sb{1}$.
\par
In fact, the occurrence of different hopping coefficients in
Eqs.~(\ref{eq:sh22})--(\ref{eq:aav12}) (ref.~%
\cite{Pl95},
reported the values $K\sb{22} \simeq -0.477$, $K\sb{11} \simeq -0.887$,
$K\sb{12} = K\sb{21} \simeq 0.834$) points to the existence of three
\emph{asymmetric} processes depending on the initial and the final energy
subbands connected by the higher order correlation functions
$\langle X\sb{i}\sp{02}X\sb{j}\sp{20}\rangle$ or
$\langle X\sb{i}\sp{02}N\sb{j}\rangle$ respectively.
\par
Finally, the present analysis strengthens the discussion in
\cite{Pl03}
concerning the unreliability of the approach of reference
\cite{SMP00}
towards the derivation of a GMFA expression for
 $\langle X\sb{i}\sp{02}N\sb{j}\rangle$ based on the
use of the Roth decoupling procedure which uncouples the Hubbard operators
at a same site,
$$
  \langle X\sb{i}\sp{02}N\sb{j}\rangle =
  \langle X\sb{i}\sp{0\sigma}X\sb{i}\sp{\sigma2}N\sb{j}\rangle =
  \langle c\sb{i\sigma} c\sb{i\bar\sigma}N\sb{j}\rangle \rightarrow
  \langle\langle c\sb{i\sigma}(t)\vert
        c\sb{i\bar\sigma}(t')N\sb{j}(t')\rangle\rangle .
$$
Therefore, the consequences following from this decoupling, namely
that special interstitial excitations ("cexons") should appear and play
an important role in the occurrence of superconductivity in cuprates, are
artifacts of the procedure without actual physical meaning.
\section{Conclusions}
We reported a method for evaluating two higher order correlation functions
describing respectively the singlet hopping and the
superconductivity pairing within the two-time Green function approach
to the solution of the two-band singlet-hole
Hubbard model~(\ref{eq:H}) considered by Plakida et al.~%
\cite{Pl95,Pl03}
for the description of the physical properties of the high-$T\sb{c}$
superconductivity in cuprates.
Both in hole-doped and electron-doped cuprates, the
dominant contributions to the two processes are found to be second order
effects described by GMFA correlation functions.
\par
The derived results are rigorously established. For the two discussed
processes, they rest on the occurrence of specific \emph{invariant} classes
of Green functions with respect to the operators of the Hubbard $1$-forms
entering the equation of motion~(\ref{eq:GFro}), allowing therefrom powers
series expansions in terms of the small Wannier overlap coefficients
$\nu\sb{ij}$.
\par
The singularity coming at the Fermi level from the thermodynamic
factor $(1-e\sp{-\beta\omega})\sp{-1}$ is canceled by a corresponding
singularity coming from the pole of the Green function at the Fermi level.
From the point of view of the properties of the functions of complex
variables, at $\omega=0$ there arises a pole of the second order yielding
the finite second order contributions found in the previous section.
\par
We have to remark that the existence of a non-vanishing commutator
$[X\sb{i}\sp{02}, H]$ is essential.
Since $[N\sb{i}, H] = [X\sb{i}\sp{\sigma\bar\sigma}, H] = 0$, the method
described in this paper cannot be used to the GMFA evaluation of the
charge-charge and spin-spin correlation functions.
\par
{\small {\bf Acknowledgments.} We are deeply indebted to Prof. N.M. Plakida
for introducing us to the Hubbard model, for many illuminating discussions,
and for critical reading of the manuscript. Useful discussion with
Dr.~A.M.~Chervyakov are gratefully acknowledged.
This research was partially financed within the CERES Contract
4-158/15.11.2004 in IFIN-HH and theme 09-6-1060-2005/2007 in LIT-JINR.
The authors acknowledge the financial support received
within the "Hulubei-Meshcheryakov" Programme, JINR order
No. 726/06.12.2005 for participation to MMCP 2006 Conference.}

\vfil\eject
\def\endpage{\hfill\eject}

\end{document}